# Kinetics of $^4$He gas sorption by fullerite $C_{60}$. Quantum effects

A.V. Dolbin, V.B. Esel'son, V.G. Gavrilko, V.G. Manzhelii, N.A.Vinnikov, S.N. Popov

B. Verkin Institute for Low Temperature Physics & Engineering NASU, Kharkov 61103, Ukraine

Electronic address:  dolbin@ilt.kharkov.ua

## Abstract

The kinetics of helium gas sorption by a $C_{60}$ powder and subsequent desorption of the $^4$He impurity from the saturated powder has been investigated in the temperature interval T = 2—292 K. Evidence is obtained that supports the existence of two stages in the temperature dependences of sorption and desorption. The stages account for the different times taken by helium to occupy the octahedral and tetrahedral interstices in the $C_{60}$ lattice. The characteristic times of sorption and desorption coincide. It is found that the temperature dependences of the characteristic times of occupying the octahedral and tetrahedral interstices are nonmonotonic. As the temperature is lowered from 292 K to 79.3 K, the characteristic times increase, which indicates a predominance of thermally activated diffusion of helium in $C_{60}$. On a further decrease to T = 10 K the characteristic times reduce over an order of magnitude. Below 8 K the characteristic times of sorption and desorption are temperature-independent**.** This suggests a tunnel character of $^4$He diffusion in $C_{60}$.

## Introduction

Since the discovery of fullerenes in 1985 [1], carbon nanomaterials have been attracting continuous attention. Among the fullerenes known, $C_{60}$ has been investigated most intensively. $C_{60}$ molecules form a molecular crystal - fullerite which has a FCC lattice above 260 K. In the fullerite crystal $C_{60}$ molecules are orientationally disordered and perform a weakly hindered rotation.  At T ~ 260 K $C_{60}$ undergoes a phase transition to a partly orientationally-ordered phase having a simple cubic lattice. On a further decrease in the temperature  the orientations of the molecules grow quenched at T ~ 90 K and an orientational glass is formed in which the $C_{60}$ molecules have no long-range orientation order. The crystal lattice of $C_{60}$ has large enough interstitial voids which can be occupied by relatively small molecules of other substances. There are one octahedral and two tetrahedral voids per molecule within the $C_{60}$ lattice.

Occupation of these voids by impurities can lead to the formation of systems with qualitatively new properties. Such changes are particularly pronounced at low temperatures at which quantum effects appear. In particular, when the interstitial cavities of a crystal $C_{60}$ lattice are filled with atoms of alkali metals, such a solution can change into the superconducting state at temperatures below 33K [2]. The occupation of a $C_{60}$ lattice by gas atoms and molecules provokes phase transformations between different orientational glasses of $C_{60}$ near the boiling temperature of liquid helium [3-9].

Quantum effects can be caused by the tunnel motion of the impurity atoms or molecules in $C_{60}$. The probability of the tunnel motion of impurities increases as the impurity – $C_{60}$ lattice interaction and the mass and size of the impurity atom/molecule decrease. $^4$He-$C_{60}$ solutions therefore seem to be most promising for detecting the tunnel motion of impurity molecules. The kinetics of sorption and desorption is essentially dependent on the impurity diffusion and must be sensitive to the mechanism of impurity (He atoms) travel in solid $C_{60}$.

To our knowledge, the kinetics of helium sorption in fullerite $C_{60}$ has been investigated by two research groups. The investigation of the effect of He atom intercalation upon the lattice parameter of $C_{60}$ by the X-ray diffraction method [10, 11] shows that the saturation of $C_{60}$ with $^4$He proceeds in two stages. According to the authors' conclusion, the impurity fills first the octahedral and then the tetrahedral subsystems of voids in the $C_{60}$ crystal. Note that the latter subsystem is filled about an order of magnitude slower than the octahedral one. The kinetics of $^4$He and $^3$He sorption in a 80% $C_{60}$-20% $C_{70}$ fullerite mixture was investigated at T = 77 K and 300 K [12]. The authors [12] described the time dependence of sorption with a single exponent and observed a slight decrease in the sorption time as the temperature changed from 300 K to 77 K.

In this study the temperature dependence of the kinetics of sorption and subsequent desorption of $^4$He from a $C_{60}$ powder was investigated in the interval 2—292 K using the technique of measuring the time dependence of the pressure of the $^4$He gas that was in contact with the $C_{60}$ powder within a closed volume.

## 1. Experimental technique

The kinetics of $^4$He sorption and desorption in the $^4$He-$C_{60}$ system was investigated on a laboratory bench, its design and operation being detailed elsewhere [13]. The used $C_{60}$ powder of mass 515.2 mg had grains about 1 μm in size. Its purity was 99.99 wt. %. To remove possible gas impurities and moisture, the powder was evacuated before measurement for 72 hours at T ~ 450 $^o$C. Then during a short time (~30 min. in the air atmosphere) the powder was transferred to the measuring cell and evacuated again at room temperature for 48 hours.

The kinetics of $^4$He sorption and desorption was investigated at T = 292 K, then at T = 79.3 K and in the interval 2—10 K. At T = 292 K and 79.3 K $C_{60}$ was saturated under the $^4$He pressure of 760 Torr. In the interval 2—10 K the pressures of the $^4$He gas were 2 and 10 Torr. Since at T = 2—10 K the pressure of the He gas in the measuring cell was much lower than the saturated vapor pressure of helium (23.8 Torr at T = 2 K [14]), this prevented condensation of the $^4$He vapor and the formation of $^4$He films on the surface of the powder grains and the cell walls.

The pressure variations in the closed volume of the measuring cell were measured continuously during saturation using a capacitive pressure transducer (MKS "Baratron") permitting us to measure low pressures with the error $1*10^{-3}$ Torr. On reaching the equilibrium pressure, the $^4$He gas was removed fast (~1 min) from the measuring cell and the cell was sealed again. The pressure variations were measured in the process of $^4$He desorption from the powder. When the investigation of desorption was completed and the pressure in the cell with the powder became equilibrium, the $^4$He impurity that might still remain in the powder was removed through evacuation at room temperature. The pressure control in the measuring cell after evacuation showed a practically total absence of $^4$He desorption from the powder in the vacuum $10^{-3}$ Torr. After removing helium, the powder was cooled down to the subsequent temperature of measurement and the process of saturation was repeated. The data acquisition system of the laboratory bench allowed recording the gas pressure at an interval of 0.2 s and thus enabled a reliable registration of high-speed pressure variations in the measuring cell.

## 2. Results and discussion

In this study we have obtained evidence that supports the previous results [10, 11] on a two-stage process of saturation of $C_{60}$ with the $^4$He impurity. The experimental dependence of pressure variations in the course of He sorption or desorption (see Fig.1) was approximated by a sum of two exponential functions having different characteristic times ($\tau_1$, $\tau_2$).

$$P = A \cdot ((1 - \exp(-t/\tau_1)) + (1 - \exp(-t/\tau_2))) + C \qquad (1)$$

The parameters $\tau_1$, $\tau_2$, A and C were obtained by fitting Eq.(1) to the experimental results.

The conclusions of Refs. [10, 11] suggest that the exponents $\tau_1$ and $\tau_2$ correspond to the characteristic times during which the He atoms can occupy the octahedral and tetrahedral subsystems of the interstitial sites in the $C_{60}$ lattice. The characteristic times of occupying the octahedral interstices ($\tau_1$) are one or two orders of magnitude shorter than the characteristic times taken to occupy the considerably smaller tetrahedral voids ($\tau_2$).

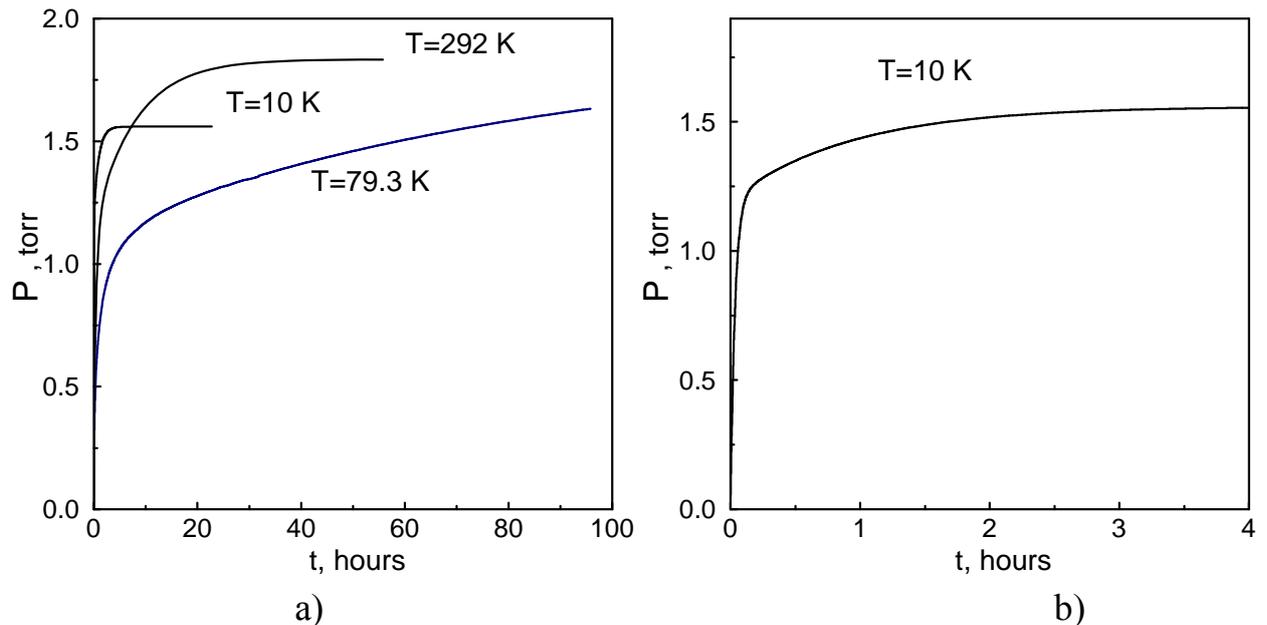

Fig.1. Experimental curves of pressure variations in the course of He desorption from a $C_{60}$ powder, the temperatures of the sample being a) 292 K, 79.3 K, 10 K; b) 10 K.

Note that the time dependences of pressure variations measured on $^4$He sorption and desorption at the same values of the powder temperature are qualitatively similar and their characteristic times coincide within the accuracy of the measuring technique. Besides, at T = 2—10 K the characteristic times $\tau_1$ and $\tau_2$ were independent of the starting pressure of the $^4$He gas in the measuring cell. The temperature dependences of the characteristic times of $^4$He sorption by $C_{60}$ powder and the proper times of $^4$He gas thermalization in the measuring system are illustrated in Fig. 2. The thermalization times were estimated in the absence of $C_{60}$ powder in the measuring cell. The characteristic times $\tau_1$ and $\tau_2$ exceed the proper times of $^4$He thermalization at least by an order of magnitude in the whole interval of the measurement temperatures. Thus, the proper times of $^4$He gas thermalization in the measuring cell have practically no effect on the temperature dependences $\tau_1(T)$ and $\tau_2(T)$.

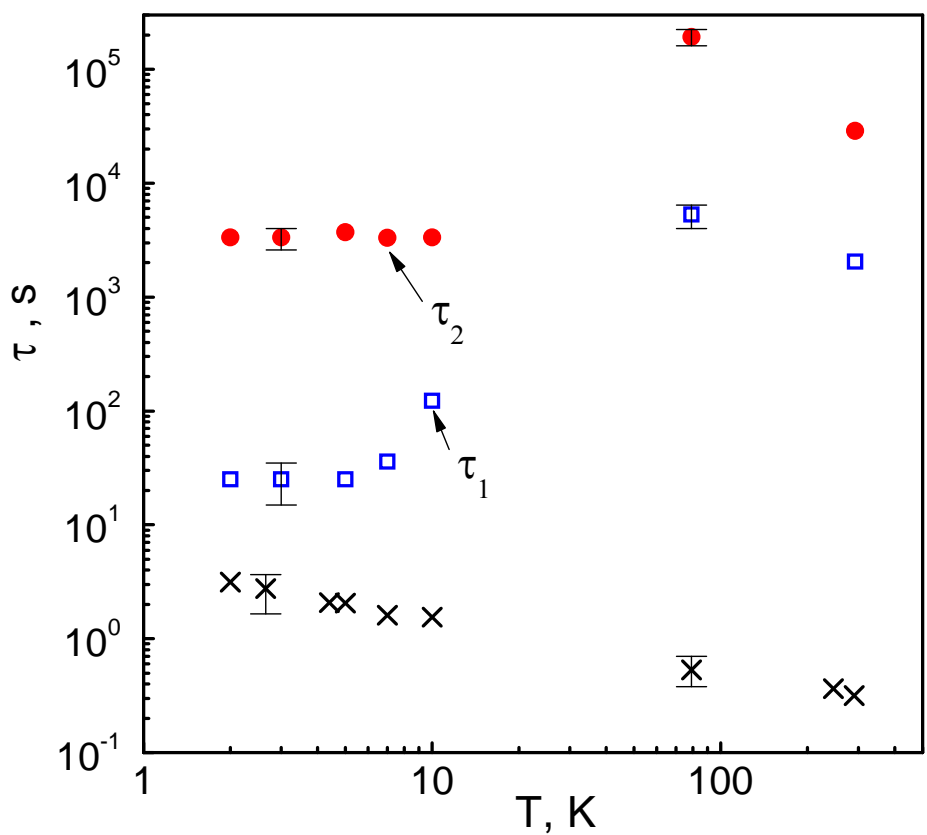

Fig.2. The temperature dependence of the characteristic times of $^4$He sorption by fullerite $C_{60}$ (circles - $\tau_2$, squares - $\tau_1$, x – proper times of He thermalization in the measuring system).

As the temperature is lowered from 292 K to 79.3 K, the characteristic times of $^4$He sorption by $C_{60}$ increase, which corresponds to the thermally activated mechanism of diffusion. However, on a further drop of the powder temperature (to 10 K and lower) the characteristic times of sorption decrease sharply. Besides, below 8 K the characteristic times are temperature independent. This leads us to assume that below T = 10 K tunneling of $^4$He atoms becomes a dominant process in fullerite which determines the rate of $^4$He sorption (desorption) in $C_{60}$.

It should be noted that the quantity of helium sorbed by the $C_{60}$ powder at T = 2—10K under the $^4$He gas pressure 10 Torr does not exceed 2 mol. %. This agrees with the conclusions of experimental investigations in which the limiting solubility of helium in fullerite in the temperature interval 1.50—1.68 K was no more than 5% [15, 16]. On the other hand, our estimates of the $^4$He concentration are over an order of magnitude higher than the amount of helium that could be sorbed at the grain boundaries in $C_{60}$ polycrystals with a grain size of ~1 μm.

## Conclusions

The performed investigations of the kinetics of $^4$He sorption by a $C_{60}$ powder and the subsequent $^4$He desorption from the saturated powder show that the characteristic times of occupation of the octahedral and tetrahedral interstitial voids in the $C_{60}$ lattice exhibit a nonmonotonic dependence on temperature. As the temperature of the sample lowers from room temperature to 79.3 K, the characteristic times of $^4$He sorption by $C_{60}$ increase, which corresponds to thermally activated $^4$He diffusion in $C_{60}$. A further decrease in the temperature of the $C_{60}$ powder causes a sharp reduction of the characteristic times in the interval 2—10 K. Besides, at T = 2—8 K the characteristic times are independent of temperature. This suggests that below 10 K tunneling of He atoms becomes dominant in $C_{60}$ and determines the rate of $^4$He sorption.

The authors are indebted to Prof. L.A. Pastur for fruitful discussions.

**Bibliography**

1. H.W. Kroto, J.R. Heath, S.C. O'Brien, et. al., *Nature* **318**, 162 (1985)
2. A.F. Hebard *Annu. Rev. Mater. Sci.*, **23**, 159 (1993)
3. A.N. Aleksandrovskii, A. S. Bakai, A. V. Dolbin, G. E. Gadd, V. B. Esel'son, V. G. Gavrilko, V.G.Manzhelii, B. Sundqvist, and B.G. Udovidchenko, *Fiz. Nizk. Temp.* **29**, 432 (2003) [*Low Temp. Phys.* **29**, 324 (2003)].
4. A.N. Aleksandrovskii, A.S Bakai, D. Cassidy, A.V. Dolbin, V.B. Esel'son, G.E. Gadd, V.G. Gavrilko, V.G. Manzhelii, S. Moricca, B. Sundqvist, *Fiz. Nizk. Temp.* **31**, 565, (2005) [*Low Temp. Phys.* **31**, (2005)].
5. V.G. Manzhelii, A.V. Dolbin, V.B. Esel`son, V.G. Gavrilko, D. Cassidy, G.E. Gadd, S. Moricca, and B. Sundqvist, *Fiz. Nizk. Temp.* **32**, 913, (2006) [*Low Temp. Phys.* **32**, 695 (2006)].
6. N.A.Vinnikov, V.G. Gavrilko, A.V. Dolbin, V.B. Esel`son, V.G. Manzhelii, B. Sundqvist, *Fiz. Nizk. Temp.* **33**, 618 (2007) [*Low Temp. Phys.* **33**, 465 (2007)].
7. A.V. Dolbin, V.B. Esel`son, V.G. Gavrilko,V.G. Manzhelii, N.A.Vinnikov, G.E. Gadd, S. Moricca, D. Cassidy, B. Sundqvist, *Fiz. Nizk. Temp.* **33**, 1401 (2007) [*Low Temp. Phys.* **33**, 1068 (2007)].

8. A.V. Dolbin, V.B. Esel`son, V.G. Gavrilko, V.G. Manzhelii, N.A. Vinnikov, G.E. Gadd, S. Moricca, D. Cassidy, B. Sundqvist, *Fiz. Nizk. Temp.* **34**, 470 (2008) [*Low Temp. Phys.* **34**, 465 (2008)].
9. A. V. Dolbin, V. B. Esel'son, V. G. Gavrilko, V. G. Manzhelii, S. N. Popov, and N. A. Vinnikov, N. I. Danilenko, B. Sundqvist, *Fiz. Nizk. Temp.* **35**, 299 (2009) [*Low Temp. Phys.* **35**, 226 (2009)]
10. Yu. E. Stetsenko, I. V. Legchenkova, K. A. Yagotintsev, A. I. Prokhvatilov, and M. A. Strzhemechny, *Fiz. Nizk. Temp.* **29**, 597–602 (2003) [*Low Temp. Phys.* **29**, 445, (2003)]
11. K. A. Yagotintsev, M. A. Strzhemechny, Yu. E. Stetsenko, I. V. Legchenkova, A. I. Prokhvatilov. *Physica B* **381,** 224-232 (2006).
12. C. P. Chen,; S. Mehta, L. P. Fu, A. Petrou, F. M. Gasparini, A. Hebard, *Phys. Rev. Lett.* **71**, pp.739-742, (1993).
13. A. V. Dolbin, V. B. Esel'son, V. G. Gavrilko, V. G. Manzhelii, N. A. Vinnikov, S. N. Popov, N. I. Danilenko and B. Sundqvist, *Fiz. Nizk. Temp.* **35**, 613 (2009) [*Low Temp. Phys.* **35**, 484 (2009)]
14. F. Brickwedde e.a., *Journ.Res.Nat Bureau Stand.* 64A, pp.1-17 (1960)
15. W. Teizer, R.B. Hallock, and A.F Hebard, *J.Low Temp Phys.* **109**, 243 (1997)
16. W. Teizer, R.B. Hallock, Q.M. Hudspeth; A.F. Hebard, *J. Low Temp. Phys.*, **113**, pp. 453-458 (1998)